# Ce-doped $Gd_3Al_2Ga_3O_{12}$ scintillator for compact, effective and high time resolution detector of the fast neutrons

Mikhail Korzhik[a,b]*, Kai-Thomas Brinkmann[c], Georgy Dosovitskiy[a], Valery Dormenev[c], Andrei Fedorov[a,b], Dmitry Kozlov[b], Vitaly Mechinsky[a,b], Hans-Georg Zaunick[c]

[a] *National Research Center "Kurchatov Institute", Moscow, Akad. Kurchatov square 1, 123182 Russia*
[b] *Institute for Nuclear Problems of Belarus State University, Minsk, Bobruiskaya str. 11, 220030, Belarus*
[c] *2nd Physics Institute, Justus Liebig University, Giessen, Heinrich-Buff-Ring 16, 35392, Germany*
* E-mail: mikhail.korjik@cern.ch

ABSTRACT: Gadolinium-aluminum-gallium garnet $Gd_3Al_2Ga_3O_{12}$:Ce scintillator is demonstrated to be an excellent scintillation material for detector of fast neutrons for the first time. Moreover, its application allows to obtain different responses to neutrons depending on their energy spectrum. This is achieved because the material, firstly, has high content of Gd, which absorbs neutrons with following emission of $\gamma$-quanta, and, secondly, detects this radiation efficiently thanks to high stopping power, fast scintillation kinetics, high scintillation light yield and good energy resolution. It was shown by simulation that several characteristic regions could be distinguished in $\gamma$-quanta pulse height spectra acquired with GAGG:Ce crystal under neutron irradiation, with energies nearly 90 and 190, and 511 keV; all these lines have different relative intensities depending on incident neutrons kinetic energy. This was confirmed by followed measurements with Am-Be neutron source. Scintillator shows coincidence time resolution better than 200ps for γ-quanta of 511 keV energy, what makes it the material of choice to separate neutrons by time-of-flight technique. Combining of detecting properties creates prospects to construct compact and fast neutron detector, capable of detection of each particle individually and registration of its specific response. This possibility could bring benefits to neutron-based examination techniques, e.g. utilizing time of flight methods and new capabilities of accelerator based bright neutron sources.

## Contents

## 1 Introduction

Nowadays, an unprecedented spinoff of the achievements of the experimental high energy physics to an non-destructive inspection, safety systems is carried on. Particularly, this is devoted to a wide range of physical methods and equipment for detection of different types of ionizing radiation. Among them, neutron counting with detection of each particle individually with registering its specific response in a wide energy range is one of most challenging tasks. One of its applications is in national security, in radiation portal monitors for cargo scanning. Neutron counting allows to distinguish nuclear materials, $^{233}$U, $^{235}$U, $^{237}$Np, Pu*, from medical and industrial radionuclides and naturally occurring radioactive materials [1-2]. So called active interrogation is based on irradiation of an object under analysis by probing beams, initiating neutron-induced reactions, such as elastic scattering, (n,γ), (n,p), (n,n') and neutron activation; it has much better selectivity, but requires irradiation of all analyzed objects [4]. Meanwhile, detection tools for the presence of nuclear and radionuclide materials with sufficient selectivity could provide protection by non-intrusive means.

Cold neutrons detection was found to be a powerful tool to construct neutron microscope, sophisticated tool for non-destructive inner scanning of the objects with a micron resolution [5,6].

Another fast-growing segment of measurements, applying neutron detection is a planetary nuclear spectroscopy, which utilizes the measurements of γ-rays, X-rays and neutrons emitted from the planet surface in order to locate major chemical elements and to make maps of hydrogen- and carbon-containing compounds distribution [7].

No less challenging domain of measurements, exploiting neutrons, is exploration of earth formation, particularly to characterize reservoirs of hydro-carbons [8]. Typically, a borehole is scanned, and formation is exposed to a $^{252}$Cf source or a 14 MeV neutron gun and induced radiation is detected. Every scanning is an expensive procedure, so there is a need to collect maximum information from one. Separation of epithermal and fast neutrons could increase additional information, however, it is hardly achievable using commercially available neutron-sensitive inorganic scintillation glasses and crystals [9 – 11].

The other applications, which could benefit from possibility of new means of fast neutron detection are such examination techniques as neutron radiography utilizing time of flight methods and new capabilities of accelerator based bright fast neutron sources [12-14]. Here exceptionally interesting would be a combining in one detector of the fast neutron detection and creation of the time stamp, for which purpose high energy X-rays, accompanying producing of the neutrons are utilized.

The technique of thermal and epithermal neutron routine measurements is often based on the interaction with several isotopes with high neutron capture cross section: $^3$He, $^6$Li, $^{10}$B. Concentration of such interacting nuclei in a unit volume of the detector determines the sensitivity of a method.

$^3$He proportional gas counters with moderate pressure have acceptable sensitivity in the energy region of thermal and epithermal neutrons [15]. Increase of gas pressure in a counter up to and above 1.5-2 MPa leads to a significant reduction of its detector properties, so the effective absorbing nuclei concentration is limited to ~5*10$^{20}$ atoms cm$^{-3}$.

Scintillation method of neutron detection is based on the absorption of neutrons by nuclei with high neutron interaction cross-sections and registration of secondary radiation, gamma-quanta or charged particles, by a scintillation detector. Higher concentration of neutron absorbing nuclei is typically achieved in solid state scintillators made from isotopically enriched raw materials, e.g. up to 1.5*10$^{22}$ atoms cm$^{-3}$ for $^6$Li-based glasses. It leads to higher available detector efficiency, allowing construction of more compact detectors. Moreover, scintillation signal readout using silicon photomultipliers (SiPMs), rapidly developing past ten years, allows relatively compact and low power-consuming back-end electronics and power supply compared to various gas-based detectors [16]. Combination of nuclei with high neutron capture cross-section and perfect scintillation properties in a single detection media generally could provide better possible detection sensitivity due to effective neutron absorption and complete secondary radiation registration.

The major channel of reactions of lithium with thermal neutrons is $^6$Li(n,α)T with total energy release ~ 4.8 MeV; alpha-particles and tritons, which are born in this reaction, have short passes and great probability to be absorbed in the detector media, causing scintillation. Due to this, $^6$Li based scintillators are widely used in neutron scintillation detection. Li containing halide single crystals, such as elpasolites, e.g. Cs$_2$LiYCl$_6$ :Ce, possess high light yield, up to 100 000 photons per neutron, but are very difficult in production. On the contrary, lithium inorganic glasses and glass ceramics are technologically attractive, but have a significantly lower yield of scintillations of about 6000-8000 photons per neutron [17, 18].

Boron isotope absorbs neutrons according to the reaction $^{10}$B(n,α)$^7$Li with energy release about 2.8 MeV, of which 0.48 MeV are gamma-quanta. $^{10}$B is attractive due to its lower cost for isotopically enriched compounds compared to lithium. Boron-based inorganic scintillation materials with acceptable scintillation yield have not been discovered yet [18]. It is typically used as a sputtered layer of B or $B_4C$ on top of some sort of detector, capable of detecting secondary particles [19].

A common drawback of all light nuclei capturing neutrons is the monotonic decrease of the neutron absorption cross section, falling by *1/v* law with increase of the neutron speed *v* in corresponding neutron energy range up to about $10^5$ eV, which makes the direct registration of such neutrons less effective. It is possible to use neutron moderators based on graphite, plastics etc. with subsequent registration of thermalized neutrons, but such design increases dimensions of the detectors significantly and limits sensitivity due to unavoidable neutron scattering and absorption.

Natural mixture of gadolinium isotopes has the largest cross section of thermal neutrons absorption ($E_n$ = 0.0253 eV) among natural isotope mixtures of all the stable elements, which is equal to ~49,000 barns [20]. This far exceeds that of $^3$He, $^6$Li, or $^{10}$B nuclei, and unlike $^3$He and $^6$Li, Gd does not require expensive isotopic enrichment to be an effective neutron absorber. Absorbing properties of different neutron detection media are compared in Table 1. Being non-enriched, only presence of $^{157}$Gd isotope in natural mixture makes GAGG crystal superior to enriched materials.

| Detecting compound | n absorbing isotope | | | n total absorption cross-section [a] for the isotope, b | | Absorbing layer thickness L, mm | Absorption efficiency (fraction of absorbed neutrons) for the layer L, % | |
|---|---|---|---|---|---|---|---|---|
| | Type | Enrichment | Content, at./cm$^3$ | $E_n$ = 0.025 eV | $E_n$ = 1 MeV | | $E_n$ = 0.025 eV | $E_n$ = 1 MeV |
| $^3$He gas counter, gas pressure 16 atm. | $^3$He | ~99.9% $^3$He | 4,3*10$^{20}$ | 5319.6 | 2.8727 | 20[b] | 98.96 | 0.25 |
| Li$_2$O-SiO$_2$:Ce$^{3+}$ glass | $^6$Li | Natural Li | 1,5*10$^{21}$ | 955.4 | 1.2504 | 5 | 93.87 | 0.09 |
| | | $^6$Li, 90% | 1,7*10$^{22}$ | | | 5 | 99.99 | 1.05 |
| Gd$_3$Al$_2$Ga$_3$O$_{12}$:Ce$^{3+}$ crystal | $^{155}$Gd, $^{157}$Gd, others | Natural Gd | 1,3*10$^{22}$ | 46095.4 | 5.0960 | 5 | 99.99(99) | 3.25 |

**Table 1**. Comparison of the absorbing properties of different neutron detection media. Total absorption cross-section -a [20]; b-typical diameter of a helium tube counter.

Gadolinium isotopes are heavy elements, so characteristic for them broad zone of resonances increases the neutron absorption efficiency for neutron energies from 1 eV up to 10 keV approximately. This makes Gd an attractive candidate for absorber in a scintillator for neutron detection in a wide energy range.

Moreover, starting from about 55 keV threshold of the neutron energy, the process of the neutron inelastic scattering is accompanied by the gamma-quanta emission, forming multiple soft lines in the resulting gamma-quanta spectrum [20]. This gives a potential possibility to create a Gd-based scintillator, sensitive to energy of incident neutrons. The currently used methods of neutron spectroscopy, based on Bonner spheres, analysis of angle and energy distribution of recoils, time of flight measurements etc., could provide necessary results, but are rather bulky, and hardly could be used in compact, out-of-the-laboratory applications [15].

Figure 1 shows comparison of total neutron interaction cross sections of $^6$Li and natural mixture of Gd isotopes in a wide range of neutron energies, from $10^{-5}$ eV to 20 MeV [20]. This data is in full agreement with results of absorption efficiency calculations, obtained by us using GEANT4 simulation software package [21].

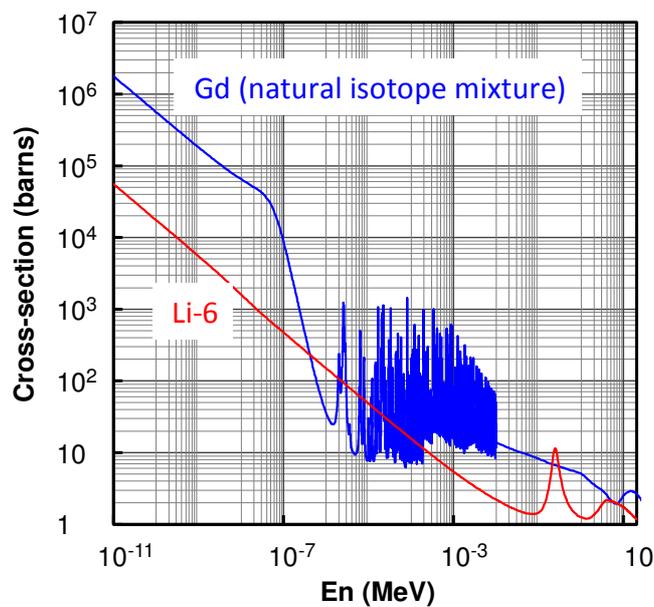

**Figure 1**. Neutron total cross sections of $^6$Li isotope and of natural mixture of Gd isotopes [20].

There are some known implementations of Gd-based neutron detectors, aimed at the registration of thermal neutrons. Reeder proposed to use a slice of crystal scintillator $Gd_2SiO_5$:Ce, surrounded by a plastic scintillator, operating together as a total absorption spectrometer of gamma rays [22]. The method of neutron counting proposed by Galunov et al. is based on a composite scintillation material consisting of $Gd_2SiO_5$:Ce or $Gd_2Si_2O_7$:Ce crystal pieces and silicone binder [23]. Unavoidable disadvantages of the methods utilizing relatively light composite materials are inability to measure emitted gamma quanta energies correctly and low detection efficiency of gamma-quanta with higher energies (in the range from 100 keV and above) emitted upon interaction of gadolinium atoms with fast neutrons. This could be overcome by use of dense and high light yield Gd-based crystalline scintillator as a detection material, which provides better efficiency and time and energy resolution of detection of secondary γ-quanta.

Here we evaluated $Gd_3Al_2Ga_3O_{12}$:Ce (GAGG:Ce) scintillation crystal [24-34] to construct a neutron-sensitive scintillation detector, capable to evaluate energy characteristics of neutron radiation. These scintillators have great potential for γ-radiation spectrometry [35]. We have discovered recently that GAGG:Ce allows to register different γ-radiation spectra on irradiation by neutrons with different energies [36]. At first, we have applied GEANT4 modeling to analyze γ-quanta spectra originating from interaction of Gd nuclei with neutrons of different energies in a wide energy ranges.. It was found, that contrary to methods, based on neutron detection with $^3$He, $^6$Li, or $^{10}$B nuclei, Gd-based detector signals contain information on the incident neutron energy. These findings were experimentally checked then. Worth to note, when photonuclear reactions are used to generate neutrons, a detection of the γ-radiation, producing these reactions, prior to a neutron flux, allows to form time a stamp with precision in a picosecond range; this introduces spectacular advantages for time-of-flight techniques of neutron detection. This gives a perspective to create a compact neutron detector with both position and energy sensitivity, and makes possible to construct new generation of neutron detectors operating in a wide energy range.

**2 Results and discussion**

**2.1 Modeling of Gd interaction with neutrons**

Yield of γ-quanta in the energy range up to 10,000 keV emitted by Gd (natural isotope mixture) upon neutron capture was modeled by GEANT4 package. Spectra were simulated with the energy step of 0.5 keV, under monochromatic neutrons with energies in range from thermal ($E_n$ = 0.0253 eV) to fast ($E_n$ = 15 MeV). Number of incident neutrons was used to be $N_n = 10^9$ for all energies. Line intensity in a certain energy range of gamma-spectrum was calculated by integrating area under the curve.

Figure 2 represents soft fraction (0 – 1000 keV) of the spectra of the emitted γ-quanta in metallic gadolinium with 2 mm thickness, irradiated with monochromatic neutrons. It is seen, that for all used neutron energies 511 keV γ-quanta line presents in spectra.

There are no prominent soft gamma-lines in the spectra for thermal neutrons ($E_n$ = 0.0253 eV) and for neutrons with $E_n$ = 10 keV. So, if registered spectrum is represented by a continuum of γ-quanta, it may indicate presence of thermal and epithermal neutrons and can be applied for their discrimination.

Gamma-lines with energies <100 keV arise in simulated spectrum for $E_n$ = 100 keV resulting from inelastic neutron scattering, which has a threshold of about $E_n$ = 55 keV. Numerous γ-quanta lines present across all the spectrum for $E_n$ = 4 MeV.

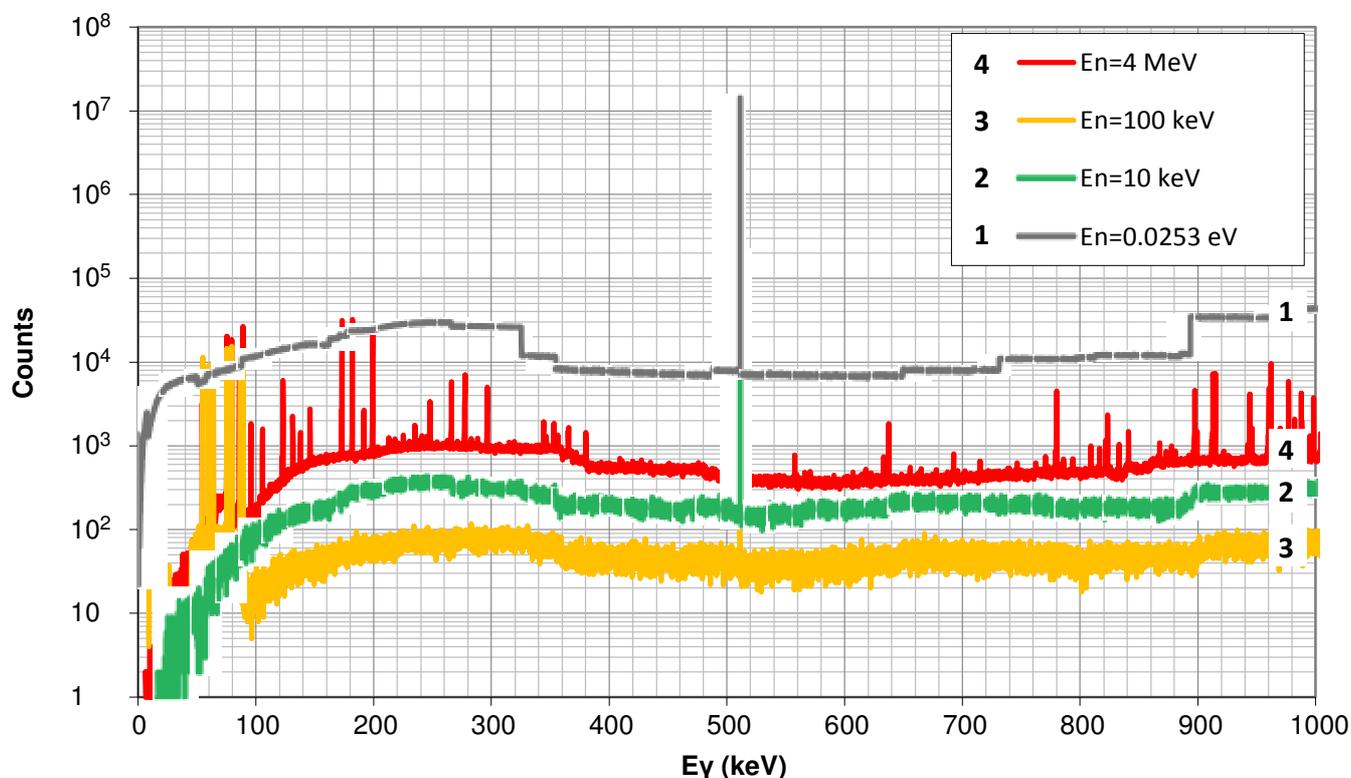

**Figure 2**. Modeled spectra of the individual energies of the emitted gamma quanta in metallic gadolinium with 2 mm thickness, irradiated with monochromatic neutrons (energies indicated on figure).

Figure 3 shows evaluated yield of γ-quanta lines (the sum of lines areas) in 40-100 keV, 100-200 keV energy ranges and for 511 keV line versus neutron energy, from $E_n$=100 keV to $E_n$=15 MeV. Each group of lines has individual dependence from neutron energy. Thus, the ratio of the integrated intensities of these groups (spectral ranges) of lines can be used to determine the spectral characteristics of the detected neutron radiation. For non-monochromatic neutron sources, an "averaged" energy (or source "signature") can be determined. The other possible way of source identification is matching the obtained values in spectral intervals with values, previously measured for a list of neutron energies.

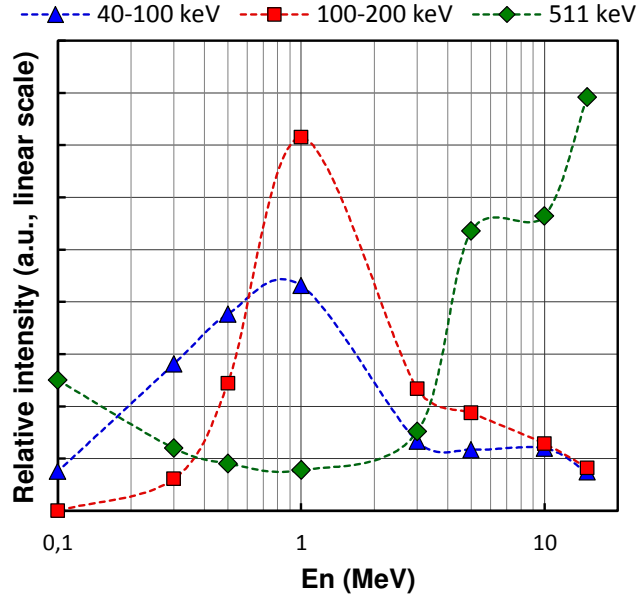

**Figure 3**. Yield of gamma lines in energy ranges 40-100 keV (blue triangles), 100-200 keV (red squares) and 511 keV (green diamonds) versus neutron energy, from $E_n$=100 keV to $E_n$=15 MeV.

For example, in order to determine a precise or "averaged" energy of the detected neutrons, the system of at least of two from the set of (1), (2), (3) equations would be solved analytically or numerically:

$$S_{40\text{-}100} = \Phi_n \times F_{40\text{-}100}(E_n) \times \lambda_{40\text{-}100} \quad (1)$$

$$S_{100\text{-}200} = \Phi_n \times F_{100\text{-}200}(E_n) \times \lambda_{100\text{-}200} \quad (2)$$

$$S_{511} = \Phi_n \times F_{511}(E_n) \times \lambda_{511} \quad (3),$$

where $\Phi_n$ is the neutron flux (or neutron radiation intensity), or neutron source activity (with appropriate calibration factor); $E_n$ is the energy or average energy of the source; $S_{40\text{-}100}$, $S_{100\text{-}200}$, $S_{511}$ − intensities of signals (lines area) detected by the detector in the energy ranges 40-100 keV, 100-200 keV and line area 511 keV respectively; $F_{40\text{-}100}(E_n)$, $F_{100\text{-}200}(E_n)$, $F_{511}(E_n)$ are the response functions of the detector in the energy ranges 40-100 keV, 100-200 keV, and line area 511 keV from the neutron energy; $\lambda_{40\text{-}100}$, $\lambda_{100\text{-}200}$, $\lambda_{511}$ − sensitivity factors of the detector to gamma quanta in the energy ranges of 40-100 keV, 100-200 keV, 511 keV.

The system (1-3) may be solved to find two unknown parameters − $\Phi_n$ and $E_n$. If necessary, the set of the equations can be extended to involve other distinguishable lines or groups of lines in wider gamma-ray energy range. In addition, a library of experimental data of the values $S_{40\text{-}100}$, $S_{100\text{-}200}$, $S_{511}$ etc. can be prepared for the expected sources. The measured values $S_{40-100}$, $S_{100-200}$, $S_{511}$ can be compared with the values from the library when detecting neutron radiation from an unknown source. Precise or good enough coincidence of the values will mean identification of the source in combination with its shielding or surrounding.

**2.2 Experiments on neutron detection**

GAGG:Ce samples used in our study were cut from single crystal boule, grown by Czochralski method from iridium crucibles by Fomos-Materials. Besides Ce activator, the crystal was co-doped with Mg and Ti [37]. Mg co-doping allows to suppress phosphorescence in the crystal whereas Ti co-doping prevents conversion of the Ce ions in crystal into a 4+ oxidation state. Light yield of the samples was ~40,000 photons /MeV, density 6.67 g cm$^{-3}$, and wavelength of scintillation spectral maximum – 520 nm. Scintillation kinetics had prevailing component with ~60 ns decay time, with approximately 80% of scintillation photons emitted within it. The rest 20% of the photons were emitted at a slower decay rate, ~150 ns.

GAGG:Ce samples with following dimensions were used for measurements: 1) 15×18×7 mm size, surface area faced to neutron source 2.70 cm$^2$, volume 1.89 cm$^3$; 2) Ø30×2 mm, surface area 7.01 cm$^2$, volume 1.41 cm$^3$.

Neutron registration measurements scheme is described below. Bialkali PMT Hamamatsu R2059 and GAGG:Ce crystal sample were placed in a 80 mm well with standard Am-Be neutron source in accordance with ISO 8529-1:2001(E). $^{241}$Am source activity was 220 GBq, estimated neutron yield equal or more than 1.3×10$^7$ neutron s$^{-1}$ and maximum of energy distribution of neutrons in the vicinity of 4MeV.

A number of optional materials were put between the neutron source and detector, as shown on measurements layout (Fig. 4). Discs of lead (14 – 40 mm) and copper (3 mm) with diameter of 80 mm were used to cut out accompanying gamma-radiation of Am-Be neutron source, in particular 4.44 MeV line, as well as a fraction of the natural gamma-ray background; copper was used specially to absorb lead characteristic X-ray radiation. Neutron absorbers based on boric acid $H_3BO_3$ with different thickness were used to transform neutron spectrum to emulate a change of neutron source "signature". Metal cadmium (2 mm thick) filter was used to cut out thermal and a fraction of epithermal neutrons in sensitivity evaluations.

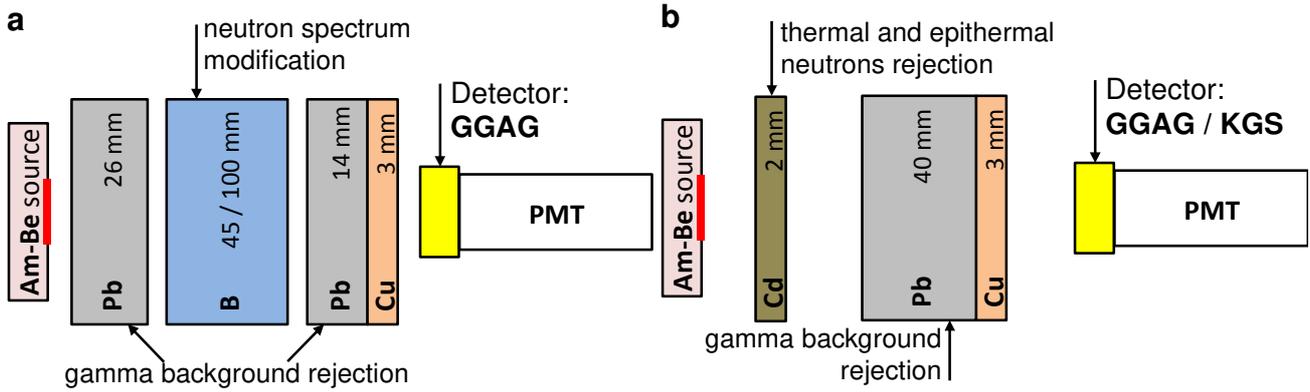

**Figure 4**. Layout of the measurements schemes: a) with boron absorber to modify neutron source energy spectrum; b) with Cd plate to cut off thermal neutrons.

Since lead and copper scatter and, in some degree, absorb neutrons from Am-Be source, its spectrum transforms and neutron flux at the detector location changes. This made a correct evaluation of GAGG:Ce detector sensitivity based on known source activity and distance to the source impossible. Instead, we have made a comparison of GAGG and $^6$Li-based scintillator sensitivities under Am-Be neutron source irradiation under the same conditions (similar distance from the source – 19 cm, and the same filter and absorber inserts).

Li-based scintillator was a KGS-3 scintillation glass ceramics with $^6$Li content close to that of well-known Saint Gobain scintillation glass GS-20, with size 18×18.5×4 mm, with surface area faced to neutron source of 3.33 cm$^2$, volume 1.33 cm$^3$ [10, 38].

Pulse height spectra of the emitted gamma-quanta have been acquired in 1000 ns time gate with charge sensitive gated ADC. The acquisition live time 300 s was used for all measurements. The light yield non-proportionality of GAGG:Ce scintillator in a low energy range and some compression of acquisition system of the energy scale above ~ 3 MeV have been taken into consideration. To improve scintillation light collection, we used Basylon® optical grease and Teflon® reflector wrapping [18].

Figure 5 shows γ-quanta spectra experimentally measured with GAGG:Ce 15×18×7 mm sample at its irradiation with neutrons from Am-Be source. Spectra were measured with combined lead (4 cm) copper (3mm) absorbers placed between detector and source, with 2 mm cadmium filter accompanying lead and copper, and without any material placed in between detector and source.

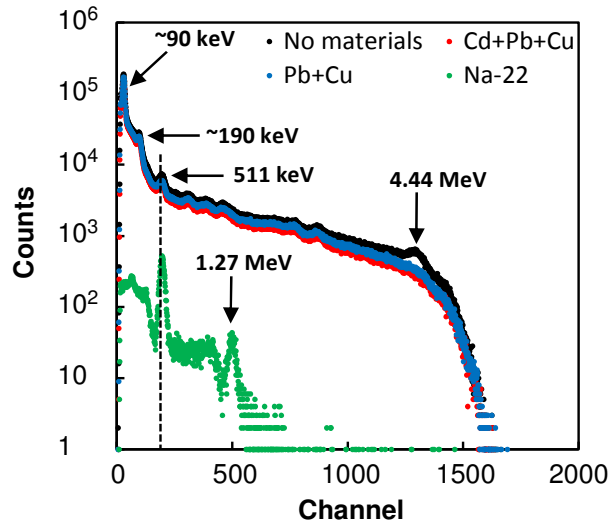

**Figure 5**. Spectra of γ-quanta measured with GAGG:Ce 15×18×7 mm scintillator under Am-Be source. Spectra, measured with lead (4 cm) and copper (3 mm) absorbers (blue), with 2 mm additional cadmium filter (red), and without any material placed in between detector and source (black). Spectrum of Na-22 source (0.511 and 1.27 MeV gammas) was recorded for energy calibration (green).

Acquired spectra exhibit all γ-quanta groups of interest, predicted in GEANT4 simulations. In addition, groups of lines with energies from 900 keV to 1200 keV and 4.44 MeV γ-quanta from Am-Be source are observed as well.

In order to check the influence of a scintillator thickness, the same measurements were repeated using GAGG:Ce Ø30×2 mm sample (figure 6). All spectral details in energy range up to 511 keV line stay in place and are clearly distinguishable.

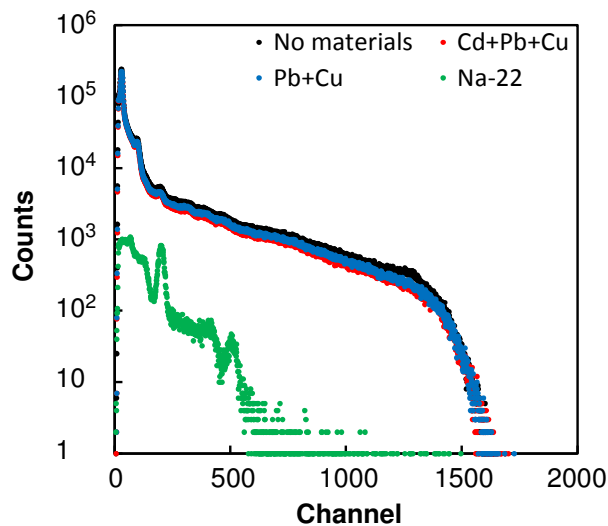

**Figure 6**. Spectra of γ-quanta measured with Ø30×2 mm$^3$ scintillaror, under Am-Be source. Spectrum of Na-22 source (0.511 and 1.27 MeV gammas) was recorded for energy calibration (green). Spectra, measured with lead (4 cm) and copper (3 mm) discs, placed between detector and source (blue), with additional 2 mm cadmium filter (red), and without any material placed in between detector and source

Figure 7 shows spectra of Am-Be source measured with GAGG:Ce 18×18×7 mm using different thickness of boric acid absorber, 4.5 cm and 10.5 cm, placed between lead discs of 2.6 cm and 1.4 cm thickness for cutting out the ~500 keV gamma line arising in boron at the neutron capture and 3 mm copper disc for cutting out the lead characteristic radiation as well.

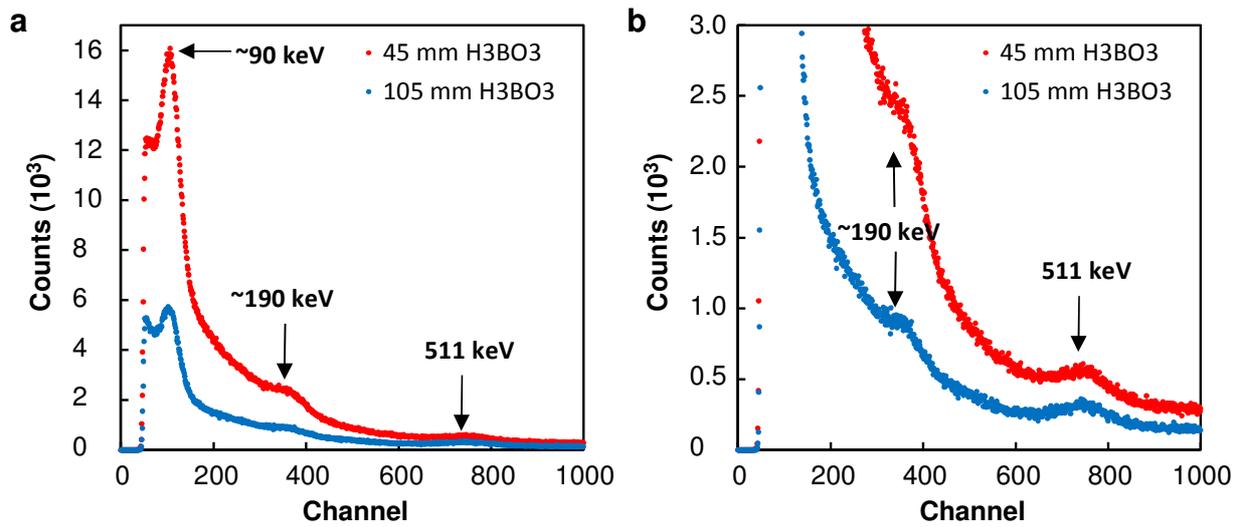

**Figure 7**. Spectra of γ-quanta measured with 15×18×7 mm scintillator under Am-Be source at different thicknesses of boron acid absorber (a). Aplitude scale is expanded at the panel (b) to demonstrate change of the lines near190 and 511 keV..

Increase of the boric acid absorber thickness from 4.5 cm to 10.5 cm results in a decrease of the area of ~90 keV line in more than 4 times, while the area of 511 keV line decreases only by a factor of 2. This is an indication of a possibility of successful solution of (1) - (3) equations in order to distinguish "signatures" of a neutron source with transformed spectra.

Amplitude spectra, recorded using $^6$Li-containing scintillation glass KGS-3 sample under neutrons from Am-Be source, contains neutron peak centered in ~800 channel (figure 8), which results from the mentioned above $^6$Li(n,α)T reaction.

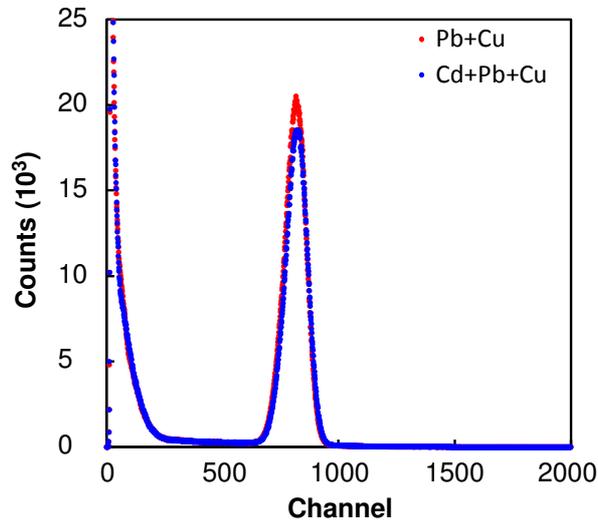

**Figure 8**. Neutron peak from Am-Be source centered in ~800 channel, recorded with $^6$Li-based KGS-3 scintillation glass.

Responses of GAGG:Ce and Li-glass to neutrons from Am-Be source are compared in table 2, which summarizes results, presented in figure 5, figure 6, figure 8.

It is seen, that count rates obtained with Cd filter and without it do not differ much; thus, we conclude that Am-Be source we have used exhibits rather low fraction of thermal and epithermal neutrons (as it should be expected). This means, in turn, that the majority of the detected neutrons keep their direction from the source. As a result, thin GAGG:Ce detector with larger area (Ø30×2 mm) gives 1.5 - 2 times better sensitivity per unit volume, than thicker detector with smaller area (15×18×7 mm). This is true for soft ~90 keV line and for channel sum over all measured gamma-spectrum, where soft gamma-quanta prevail. For 190 keV count rates become almost equal, and for 511 keV the situation tends to the opposite. All this means that for the detection of fast neutrons, the GAGG:Ce detector shape should undergo some optimization depending on required overall sensitivity and sensitivity at the specific energies.

| Detector | | Lines area (counts) | Count rate per area (count s$^{-1}$ cm$^{-2}$) | Count rate per volume (count s$^{-1}$ cm$^{-3}$) |
|---|---|---|---|---|
| KGS-3 18×18.5×4 mm | Pb+Cu | Neutron peak - 2316180 | 2318 | 5805 |
| | Cd+ Pb+Cu | Neutron peak - 2161030 | 2163 | 5416 |
| GAGG:Ce 15×18×7 mm | Pb+Cu | ~90 keV - 1255690 | 1550 | 2214 |
| | | ~190 keV - 143108 | 177 | 252 |
| | | 511 keV - 42031 | 52 | 74 |
| | | Full spectrum - 7003420 | 8646 | 12352 |
| | Cd+ Pb+Cu | ~90 keV - 1007870 | 1244 | 1778 |

| | | ~190 keV - 130794 | 161 | 231 |
| | | 511 keV - 35352 | 44 | 62 |
| | | Full spectrum - 6154690 | 7598 | 10855 |
| GAGG:Ce Ø30×2 mm | Pb+Cu | ~90 keV - 1687130 | 802 | 3988 |
| | | ~190 keV - 139606 | 66 | 330 |
| | | 511 keV - 21622 | 10 | 51 |
| | | Full spectrum - 7528330 | 3580 | 17797 |
| | Cd+ Pb+Cu | ~90 keV - 1416390 | 674 | 3348 |
| | | ~190 keV - 119538 | 57 | 283 |
| | | 511 keV - 21392 | 10 | 51 |
| | | Full spectrum - 6964720 | 3312 | 16465 |

**Table 2**. $^6$Li-based KGS glass and GAGG:Ce detector count rates measured in the same conditions at Am-Be source.

Comparison of count rates per unit volume for full gamma-spectra areas for GAGG:Ce and KGS-3 (table 2), demonstrates 2-3 times higher values for GAGG:Ce (depending on the detector geometry), which is significant advantage. However, use of the full gamma-spectrum recorded with GAGG:Ce requires careful suppression and/or accounting of natural gamma-ray background, which could be inconvenient for some practical applications. In this case expressed and easily separated spectral lines could be used, such as 90, 190, 511 keV. Yield of these lines is rather high as well, and the sum of count rates (or total area of lines) per unit volume equals 38-75% of that of $^6$Li-glass scintillator depending of scintillator shape. Additionally, GAGG:Ce neutron detector shielding from the natural background can drastically improve signal-to-noise ratio. Thus, amplitude analysis of the spectra, acquired with GAGG scintillator, can be used to discriminate neutrons of the different energy.

Another approach is based on the fast response of GAGG scintillator to γ-quanta. Coinsidence time resolution (CTR), obtained with Ce solely doped GAGG pixels, was measured to be ~480 ps [39]. Recently, tremendous progress with improvement of CTR down to 165 ps with codoped GAGG crystals and new generation of High dencity RGB0 SiPM was pointed out [40]. Table 3 shows difference in the arriving time of neutrons of nearby energies at the distance of 2 m from the source.

| Neutron energies, MeV | 0.099 and 0.1 | 0.99 and 1 | 9.9 and 10 | 99 and 100 |
|---|---|---|---|---|
| Difference in arrival at | | | | |
| the distance 2m, ps | 2000 | 500 | 260 | 50 |
| Minimal distance for time discrimination of the neutrons with nearby energies, m | 0.15 | 0.6 | 1.2 | 4.2 |

**Table 3**. Difference in arriving time of neutrons of nearby energies at the distance 2 m from the source and minimal distance from the source for their time discrimination with GAGG SiPM based detector. Application of GAGG scintillation detector allows to reduce dimensions for the neutron facilitied , operating in spectroscopic mode.

## 3 Conclusion

Modeling and experimental evaluations presented have shown, that γ-quanta spectra acquired with GAGG:Ce scintillation detector under neutron irradiation contain information on the incident neutron kinetic energy, which can be used, in particular, for neutron spectroscopy. Sensitivity of GAGG:Ce detector with moderate size in a wide range of neutron energies was found to be comparable or superior to that of $^6$Li-based glass scintillation materials, depending on method of γ-quanta counting. Moreover, GAGG:Ce does not require an enrichment with neutron sensitive isotopes, what makes it cost effective in comparison with $^3$He gas counters and $^6$Li-containing inorganic scintillators. Moreover, GAGG scintillation detector, possessing fast response to γ-quanta, allows precise time-of-flight discrimination of the fast neutrons. These findings open an opportunity to create compact neutron detectors with both position and energy sensitivity, operating in a wide energy range.

**Acknowledgements**

The work is supported by grant № 14.W03.31.0004 of Russian Federation Government. Authors are grateful to FOMOS-Crystals and NeoChem (Moscow, Russia) for the samples of GAGG scintillator for measurements.